\documentstyle[aps,prb]{revtex}

\begin{document}
\title{The hydration free energy of water}
\author{Gerhard Hummer,$^{*ab}$ Lawrence R. Pratt,$^c$ and
Angel E. Garc\'{\i}a$^a$}
\address{}
\date{May 10, 1995}
\maketitle
\begin{center}
{\em The Journal of Physical Chemistry} (in press, 1995)
\end{center}
\begin{abstract}
We study the chemical potential of water as a function of charge based
on perturbation theory. By calculating the electrostatic-energy
fluctuations of two states (fully charged and uncharged) we are able
to determine accurate values for the dependence of the chemical
potential on charge. We find identical results for the
chemical-potential difference of fully charged and uncharged water
from overlapping-histogram and acceptance-ratio methods and by
smoothly connecting the curves of direct exponential averages. Our
results agree with those of Rick and Berne ({\em J. Am. Chem. Soc.},
{\bf 1994}, {\em 116}, 3949) with respect to both the
chemical-potential difference and its dependence on the
charge coupling parameter. We observe significant deviations from
simple Gaussian-fluctuation statistics. The dependence on the coupling
parameter is not quadratic, as would be inferred from linear continuum
models of electrostatics.
\end{abstract}

\section{Introduction}
The accurate calculation of {\em thermal} properties of fluid systems
such as the free energy or the entropy using computer simulations
poses serious difficulties. Unlike mechanical quantities (e.g.,
energy, pressure) which can be expressed as averages of functions of
the phase-space coordinates, {\em thermal} quantities are related to
the partition function and thus to the volume of accessible phase
space.\cite{Frenkel:86}

We recently pursued an approach based on perturbation theory for the
chemical potential which relates the free-energy difference of two
states to fluctuations of the two equilibrium
systems.\cite{Pratt:94:a,Hummer:95:c} This has the advantage that
along with free-energy data other useful information can be extracted
from equilibrium computer simulations. Motivated by the Born-model
prediction of quadratic dependence of the free energy on ionic
charge,\cite{Born:20} we studied the free energy of simple ions in
water. Within the framework of perturbation theory, the Born model
suggests validity of a second-order treatment with the ionic charge as
a coupling parameter and, correspondingly, Gaussian statistics of the
electrostatic potential.\cite{Levy:91,Figueirido:94} Our work indeed
supported this view; but it also showed that to get {\em accurate}
values for the free energy of ionic solvation it is important to
combine fluctuation information of different charge states.

In this work we study the chemical potential of water. This quantity
plays a central role in many processes of physical chemistry and
biophysics. For instance, the chemical potential of water is the
driving force in osmotic equilibria. To study inhomogeneous aqueous
systems or mixtures such as macromolecular solutions or crystals using
grand-canonical-ensemble methods, accurate values for the chemical
potential of the particular water model are required.

We will be concerned with the electrostatic contributions to the
chemical potential. Efficient methods such as test-particle
insertion\cite{Widom:82} are available for the contributions related
to the particle volume. We will study the process of uncharging a
water molecule with the van der Waals interactions unmodified.
Fluctuation statistics and, for reference, Bennett's
overlapping-histogram and acceptance-ratio methods\cite{Bennett:76}
will be used to calculate the chemical potential. We also use a
geometrical method which smoothly connects the direct exponential
averages. We compare our results for the chemical potential with those
of Rick and Berne\cite{Rick:94} obtained from thermodynamic
integration.

We will first develop the theoretical framework. The correction for
finite system size will be discussed. As in our previous
studies,\cite{Hummer:95:c,Hummer:93} we will use Ewald lattice
summation and a generalized reaction-field (GRF) method for the
electrostatic interactions and consistently add the self-interactions
as finite-size corrections. We will then describe the computer
simulations and discuss the results.

\vfill
\hrule\vspace*{0.5\baselineskip}\noindent
{\small
$^a$ Address for Correspondence:
Theoretical Biology and Biophysics Group T-10, MS K710,
Los Alamos National Laboratory, Los Alamos, New Mexico 87545, U.S.A.
{[FAX: (505) 665-3493;
Phone: (505) 665-1923;
E-mail: {\tt hummer@t10.lanl.gov]}\\[0.5\baselineskip]
$^b$ Center for Nonlinear Studies,
Los Alamos National Laboratory, MS B258, Los Alamos, New Mexico
87545, U.S.A.\\[0.5\baselineskip]
$^c$ Theoretical Chemistry and Molecular Physics Group T-12, MS B268,
Los Alamos National Laboratory, Los Alamos, New Mexico 87545, U.S.A.}
\clearpage

\section{Calculation of chemical potentials}
\subsection{Fluctuation statistics}
Various methods are available to calculate free-energy differences
between different states of fluid systems using computer simulations
(see refs~\onlinecite{Frenkel:86,Allen:87,Levesque:92} for
reviews). Here we are concerned with determining the electrostatic
contributions to the free energy of water. The chemical-potential
difference $\Delta\!\mu^{ex}$ between a charged and an uncharged water
molecule in bulk water is calculated. Our goal is to relate
$\Delta\!\mu^{ex}$ to fluctuations in the electrostatic energy of
single water molecules.

The potential-distribution theorem for the excess chemical potential
$\mu^{ex}$ forms a convenient starting point:\cite{Widom:82}
\begin{eqnarray}
\mu^{ex}(\lambda_1)-\mu^{ex}(\lambda_0)&=&-k_BT\ln
\left\langle {\exp\left\{-\beta [u(\lambda_1)-u(\lambda_0)]\right\}}
\right\rangle_{\lambda_0}~.
\label{eq:dF}
\end{eqnarray}
$\lambda_0$ and $\lambda_1$ are coupling parameters representing the
two charge states. $\beta$ is the inverse temperature $1/k_B T$. The
thermal configuration-space average in the charge-state $\lambda$ is
denoted by $\langle \ldots \rangle_\lambda$. $u(\lambda)$ is the
charge- and configuration-dependent interaction energy of a water
molecule carrying charges $(1-\lambda)q_O$ and $(1-\lambda)q_H$ on
oxygen and hydrogen atoms, respectively. $\lambda=0$ and 1 correspond
to the fully charged and uncharged state, respectively. We will only
discuss results for three-point models of water. Extensions to water
models carrying four and more partial charges are trivial.

The energy $u(\lambda)$ contains the electrostatic interactions
$u_{el}(\lambda)$ with all other water molecules and a
self-interaction $u_s(\lambda)$,
\begin{eqnarray}
u(\lambda) & = & u_{el}(\lambda) + u_s(\lambda)~,
\end{eqnarray}
where
\begin{eqnarray}
u_{el}(\lambda) & = & ( 1 - \lambda ) \; \left[ q_O \phi_O + q_H
\left( \phi_{H_1} + \phi_{H_2} \right) \right]~.
\end{eqnarray}
$\phi_O$, $\phi_{H_1}$, and $\phi_{H_2}$ are the electrostatic
potentials at the sites of the oxygen and hydrogen atoms. The
self-interaction $u_s(\lambda)$ depends on the effective electrostatic
interaction $\varphi({\bf r})$ used in the computer simulations,
\begin{eqnarray}
u_s(\lambda) & = & \frac {1}{2} \sum_{\alpha,\beta} ( 1 - \lambda )^2
q_\alpha q_\beta \Psi(r_{\alpha\beta})~,
\label{eq:self}
\end{eqnarray}
where the double sum extends over pairs of charge sites on a water
molecule. $\Psi$ is averaged over the solid angle $\Omega$,
\begin{eqnarray}
\Psi(r_{\alpha\beta}) & = & \int_\Omega d\Omega\;\left[ \varphi({\bf
r}_{\alpha\beta})-\frac{1}{|{\bf r}|} \right]~.
\end{eqnarray}
The perturbation expression eq~\ref{eq:dF} can be used directly if the
$\lambda_1$ and $\lambda_0$ states are close. Otherwise, the average
in eq~\ref{eq:dF} is dominated by the poorly sampled tails of the
distribution of $\Delta\! u=u(\lambda_1)-u(\lambda_0)$. As in our
previous work,\cite{Pratt:94:a,Hummer:95:c} we use a perturbative
expansion. Application of the cumulant expansion\cite{Stuart:87} with
respect to $\Delta\! \lambda = \lambda_1 - \lambda_0$ yields a series
expansion for the difference in chemical potential $\Delta\! \mu^{ex}
= \mu^{ex}(\lambda_1) - \mu^{ex}(\lambda_0)$:
\begin{eqnarray}
-\beta\Delta\!\mu^{ex} & = & -\beta u_s(0) \left[ \Delta\!\lambda \left(
\Delta\!\lambda + 2\lambda_0 -2 \right) \right]
+ \sum_{n=1}^{\infty} \frac{(\beta\Delta\!\lambda)^n}{n!}
C_{n,\lambda_0} [u_{el}(0)]~.
\label{eq:cumu}
\end{eqnarray}
The cumulants $C_{n,\lambda_0}[u_{el}(0)]$ measure the fluctuations
of the electrostatic energy $u_{el}(0)$ in the state $\lambda_0$:
\begin{mathletters}
\begin{eqnarray}
C_{1,\lambda_0}[u_{el}(0)] & = & \left\langle u_{el}(0)
\right\rangle_{\lambda_0}~, \\
C_{2,\lambda_0}[u_{el}(0)] & = & \left\langle \Delta\! u_{el}^2
\right\rangle_{\lambda_0}~, \\
C_{3,\lambda_0}[u_{el}(0)] & = & \left\langle \Delta\! u_{el}^3
\right\rangle_{\lambda_0}~, \\
C_{4,\lambda_0}[u_{el}(0)] & = & \left\langle \Delta\! u_{el}^4
\right\rangle_{\lambda_0} - 3 \left\langle \Delta\! u_{el}^2
\right\rangle_{\lambda_0}^2~,
\end{eqnarray}
\label{eq:cum}
\end{mathletters}\noindent
where $\Delta\! u_{el}=u_{el}(0)-\langle u_{el}(0)
\rangle_{\lambda_0}$. For a Gaussian distribution, all cumulants for
$n\geq 3$ are zero. If the distribution is not Gaussian, the cumulant
expansion is infinite or higher-order cumulants diverge, i.e., there
does not exist an $m\geq 3$ such that $C_m\neq 0$ and $C_n=0$ for all
$n>m$.\cite{Stuart:87,Marcinkiewicz:39} Therefore, the use of only
mean and variance in eq~\ref{eq:cumu} would be exact if the underlying
distribution were Gaussian. The self-interaction $u_s$ can be seen to
correct the mean and the variance of the $u_{el}$ distributions. When
identical orders of $\Delta\!\lambda$ are added in eq~\ref{eq:cumu} we
obtain corrected cumulants $C_{n,\lambda}'$:
\begin{mathletters}
\begin{eqnarray}
C_{1,\lambda_0}' & = & C_{1,\lambda_0} - 2 ( \lambda_0 - 1 ) u_s(0)\\
C_{2,\lambda_0}' & = & C_{2,\lambda_0} - 2 u_s(0) / \beta\\
C_{m,\lambda_0}' & = & C_{m,\lambda_0} \mbox{ for } m\geq 3~.
\end{eqnarray}
\label{eq:cumcor}
\end{mathletters}\noindent
The correction of the variance is independent of the charge state
$\lambda_0$. The correction of the mean on the other hand is
proportional to the charge factor $1-\lambda_0$ and vanishes in the
uncharged case $\lambda_0=1$.

We can view eq~\ref{eq:cumu} as a
Taylor expansion of the chemical potential around $\lambda_0$, where
the cumulants contain the information about the derivatives with
respect to the coupling parameter $\lambda$.
We can combine this information about the derivatives in a $\chi^2$
fit of $\mu^{ex}(\lambda)$ to a given functional form. We will here
use polynomials $p_l$ of varying order $l$. The constant term, $a_0$,
is undetermined. The coefficients $\{a_k\}$ will be chosen to minimize
a $\chi^2$ functional,
\begin{eqnarray}
\chi^2(\{a_k\}) & = & \sum_{n=1}^{N} \sum_{m=1}^{M_n}
\left[\sigma_n^{(m)}\right]^{-2} \left[
p^{(m)}_l(\lambda_n,\{a_k\}) - d^{(m)}(\lambda_n) \right]^2~,
\label{eq:chi2}
\end{eqnarray}
where $N$ is the number of $\lambda_n$ values analyzed. $M_n$ is the
number of derivatives calculated at $\lambda_n$. $p^{(m)}_l$ is the
$m$-th derivative of the polynomial $p_l(\lambda,\{a_k\}) =
\mu^{ex}(\lambda)$ with respect to $\lambda$. $d^{(m)}$ is the
observed derivative, which is related to the cumulant $C_{m,\lambda}'$
through eqs~\ref{eq:cumu} and
\ref{eq:cumcor},
\begin{eqnarray}
d^{(m)}(\lambda) & = & - \beta^{m-1} \; C'_{m,\lambda}~.
\end{eqnarray}
The estimated statistical error $\sigma_n^{(m)}$ of
$d^{(m)}(\lambda_n)$ is assumed to be Gaussian.

We can calculate $u_{el}$ distributions at different charge states
$\lambda$ from simulations of systems with one modified water molecule
in a solution of unmodified water. For $\lambda=0$, the statistical
efficiency is greatly enhanced since we can average over all water
molecules. Using eqs~\ref{eq:cumu}, \ref{eq:cum}, and \ref{eq:chi2},
the cumulant data of the different charge states can be combined to
obtain the chemical potential as a function of the coupling parameter
$\lambda$.

\subsection{Bennett's overlapping-histogram and acceptance-ratio methods}
We also apply Bennett's method of overlapping
histograms\cite{Frenkel:86,Bennett:76,Gubbins:83} to calculate the free
energy of uncharging a water molecule. We study two states 0 and 1
with a difference in their configurational energies $\Delta\! u =
u(\lambda_1) - u(\lambda_0)$. The normalized probability densities
$p_i(x)$ ($i=0,1$) of the energy difference $\Delta\! u$ are defined
as
\begin{eqnarray}
p_i(x) & = & \left\langle \delta( \Delta\! u - x )
\right\rangle_{\lambda_i}~,
\end{eqnarray}
where $\delta(x)$ is Dirac's delta distribution.

Canonical-ensemble averages of a function $A$ of the configuration
variables in systems 0 and 1 are related through
\begin{eqnarray}
\langle A \rangle_{\lambda_1} & = &
\frac{ \langle A \exp ( - \beta \Delta\! u ) \rangle_{\lambda_0} }
     { \langle   \exp ( - \beta \Delta\! u ) \rangle_{\lambda_0} }~.
\end{eqnarray}
Using eq~\ref{eq:dF} we find:
\begin{eqnarray}
\langle A \rangle_{\lambda_1} & = &
\exp ( \beta \Delta\!\mu^{ex} ) \;
\langle A \exp ( - \beta \Delta\! u ) \rangle_{\lambda_0}~.
\label{eq:Aave}
\end{eqnarray}
Correspondingly, the probability densities $p_0$ and $p_1$ are related
through
\begin{eqnarray}
p_1(x) &=& p_0(x) \; \exp( \beta\Delta\!\mu^{ex} - \beta x )
\end{eqnarray}
with $\Delta\!\mu^{ex} = \mu^{ex}(\lambda_1) - \mu^{ex}(\lambda_0)$.
In a region of overlap between $p_1$ and $p_0$ one expects a linear
dependence of $\ln[p_1(x)/p_0(x)]$ on $x$ with slope $-\beta$ and
intercept $\beta\Delta\!\mu^{ex}$.

Applied to the case of uncharging a water molecule, we have
\begin{eqnarray}
\Delta\! u & = & -\Delta\!\lambda \; u_{el}(0)+\Delta\!\lambda \; (
\Delta\!\lambda + 2 \lambda_0 - 2 ) \; u_s(0)~.
\label{eq:dubenn}
\end{eqnarray}
The probability density $p_0$ can be calculated from histograms of the
electrostatic energy $\Delta\! u=-u_{el}(0)-u_s(0)$ of water molecules
in bulk water (state 0). To calculate $p_1$, a system has to be
studied comprising one uncharged and $N-1$ charged water molecules,
where $u_{el}$ is the fictitious electrostatic energy of the uncharged
particle obtained by turning on its charges. $p_1$ is then the
probability distribution of $\Delta\! u=-u_{el}(0)-u_s(0)$.

{}From eq~\ref{eq:Aave} we find the basic relation of Bennett's
acceptance-ratio method to calculate
$\Delta\!\mu^{ex}$:\cite{Frenkel:86,Bennett:76}
\begin{eqnarray}
\beta \Delta\!\mu^{ex} & = & \ln \frac
{ \left\langle f ( - \beta \Delta\! u + c ) \right\rangle_{\lambda_1} }
{ \left\langle f (   \beta \Delta\! u - c ) \right\rangle_{\lambda_0} }
+ c~,
\label{eq:acc}
\end{eqnarray}
where the Fermi function $f(x)=1/[1+\exp(x)]$ has been chosen to
minimize the error in the $\Delta\!\mu^{ex}$ estimate. $c$ is an
arbitrary constant. Setting it to $c=\beta\Delta\!\mu^{ex}$ minimizes
the expected error.\cite{Bennett:76} In a graphical procedure we
search for the intersection of the $\lambda_0$ and $\lambda_1$
averages in eq~\ref{eq:acc} as functions of $c$. In addition, we
expect a plateau for $\Delta\!\mu^{ex}$ calculated from
eq~\ref{eq:acc} for $c$ values close to the optimum
$c=\beta\Delta\!\mu^{ex}$. A more detailed discussion of the two
methods due to Bennett can be found in refs~\onlinecite{Frenkel:86}
and \onlinecite{Bennett:76}.

\subsection{Effective electrostatic interactions}
To avoid surface effects, fluid systems are commonly studied using
periodic boundary conditions. This results in well-known difficulties
regarding the treatment of long-range interactions which are large
even at distances comparable to the dimensions of the simulation box.
We treat the electrostatic interactions in a truly periodic format by
using Ewald lattice summation.\cite{Allen:87,Ewald:21,deLeeuw:80:a} In
the Ewald formulation, the electrostatic energy $U$ of a system
comprising $N$ water molecules carrying partial charges $q_\alpha$ can
expressed as
\begin{eqnarray}
U & = & \sum_{1\leq i<j\leq N} \sum_{\alpha=1}^{3} \sum_{\beta=1}^{3}
q_\alpha q_\beta \varphi_{EW}({\bf r}_{i_\alpha j_\beta}) +
\frac{1}{2} \sum_{i=1}^{N} \sum_{\alpha=1}^{3} \sum_{\beta=1}^{3}
q_\alpha q_\beta \psi_{EW}({\bf r}_{i_\alpha i_\beta})~.
\label{eq:EwaldWat}
\end{eqnarray}
A lattice vector ${\bf n}$ is added to the distance vector ${\bf
r}_{i_\alpha j_\beta}$ such that ${\bf r}_{i_\alpha j_\beta} = {\bf
r}_{j_\beta} - {\bf r}_{i_\alpha}+{\bf n}$ is in $[-L/2,L/2]^3$.

The self-interaction is defined as $\psi_{EW}({\bf r}) =
\varphi_{EW}({\bf r}) - 1/|{\bf r}|$. The effective Coulomb
interaction $\varphi_{EW}$ can be expressed in terms of rapidly
converging lattice sums,
\begin{eqnarray}
\varphi_{EW}({\bf r}) & = &
\sum_{\bf n}\frac{{\rm erfc}(\eta|{\bf
r}+{\bf n}|)} {|{\bf r}+{\bf n}|} + \sum_{{\bf k}\neq 0}
\frac{4\pi}{V k^2} \exp \left( -\frac{k^2}{4\eta^2} +
i{\bf k}\cdot{\bf r}\right) -\frac{\pi}{V \eta^2}~.
\label{eq:Ewald}
\end{eqnarray}
$V$ is the volume of the box, erfc is the complementary error
function, and $k=|{\bf k}|$. The two lattice sums extend over
lattice vectors ${\bf n}$ and ${\bf k}$ of real and Fourier space,
respectively.

To calculate the free energy of charging, we need expressions for the
electrostatic energy of the system when a single water molecule
carries modified charges $(1-\lambda)q_O$ and $(1-\lambda)q_H$. We
will identify the difference of the interaction energies $u(\lambda_1)
- u(\lambda_0)$ with the difference of the total energies
(eq~\ref{eq:EwaldWat}) for the two $\lambda$
values. Correspondingly, we can write for the interaction energy
$u(\lambda)$ of the modified water molecule $i=1$:
\begin{eqnarray}
u(\lambda) & = & \sum_{j=2}^{N} \sum_{\alpha=1}^{3} \sum_{\beta=1}^{3}
(1-\lambda) q_\alpha q_\beta \varphi_{EW}({\bf r}_{1_\alpha j_\beta}) +
\frac{1}{2} (1-\lambda)^2 \sum_{\alpha=1}^{3} \sum_{\beta=1}^{3}
q_\alpha q_\beta \psi_{EW}({\bf r}_{1_\alpha 1_\beta})~.
\label{eq:ulambda}
\end{eqnarray}
We identify the first and second sum with $u_{el}(\lambda)$ and
$u_s(\lambda)$, respectively. Eq~\ref{eq:ulambda} contains
self-interactions owing to the presence of lattice images. We can
average the self-interactions over the solid angle assuming
isotropy. The Ewald potential $\varphi_{EW}$ can be expanded in
spherical harmonics.\cite{Hummer:93,Slattery:80,Adams:87} In an
average of $\psi_{EW}({\bf r})$ over all orientations, the
orthogonality of the spherical harmonics yields only a constant plus
an $r^2$ term,
\begin{eqnarray}
\Psi_{EW}(r) & = & \int_\Omega d\Omega\;\psi_{EW}({\bf r})
= \xi_{EW} + \frac{2\pi}{3L^3}\;r^2~,
\label{eq:PsiEW}
\end{eqnarray}
where the constant is $\xi_{EW}=-2.837297/L$.\cite{Adams:87} From
eqs~\ref{eq:self} and \ref{eq:ulambda} we
obtain for the self-interaction
\begin{eqnarray}
u_s(\lambda) & = & - (1-\lambda)^2 \frac{2\pi}{3L^3} m^2
\end{eqnarray}
where $m$ is the dipole moment of a water molecule.

In common implementations of Ewald summation, the Fourier-space sum is
rewritten such that it scales with the product of the number of
particles times the number of ${\bf k}$ vectors.\cite{Allen:87} These
implementations have the disadvantage that electrostatic potentials at
the charge sites are not directly available. To circumvent this
problem we use an expansion of the Ewald potential in terms of
harmonic functions with cubic
symmetry\cite{Hummer:93,Slattery:80,Adams:87,vdLage:47} to calculate
the electrostatic potentials $\phi_O$, $\phi_{H_1}$, and
$\phi_{H_2}$. We include in the expansion kubic harmonics up to tenth
order. The coefficients are those previously used in the calculation
of chemical potentials of restricted-primitive-model
ions.\cite{Hummer:93} They are listed in Table \ref{tab:harm} in the
convention of Adams and Dubey.\cite{Adams:87} For the calculation of
the interaction energies in the Monte Carlo simulations, we do not
use the kubic-harmonic approximation but conventional lattice sums.

We also use a generalized reaction-field (GRF) method for the
electrostatics.\cite{Hummer:94:a,Hummer:94:e} The GRF interaction
depends only on the distance and has a cutoff length $r_c$,
\begin{eqnarray}
\varphi_{GRF}(r) & = &
\frac{1}{r} \; p(r/r_c) \; \Theta(r_c-r) \label{eq:GRF}~.
\end{eqnarray}
$\Theta$ is the Heaviside unit-step function; $p(x)$ is a polynomial:
\begin{eqnarray}
p(x) & = & ( 1 - x )^4 ( 1 +8x/5 +2x^2/5 )~.
\label{eq:polyGRF}
\end{eqnarray}
By analogy with Ewald summation (eq~\ref{eq:ulambda}), we define
the $\lambda$-dependent energy as
\begin{eqnarray}
u(\lambda) & = & \sum_{j=2}^{N} \sum_{\alpha=1}^{3} \sum_{\beta=1}^{3}
(1-\lambda) q_\alpha q_\beta \varphi_{GRF}(r_{1_\alpha j_\beta}) +
\frac{1}{2} (1-\lambda)^2 \sum_{\alpha=1}^{3} \sum_{\beta=1}^{3}
q_\alpha q_\beta \Psi_{GRF}(r_{1_\alpha 1_\beta})~.
\label{eq:ulambdaGRF}
\end{eqnarray}
The self-interaction is defined as
$\Psi_{GRF}(r)=\varphi_{GRF}(r)-1/r$. The GRF self-term for an
uncharged molecule contains a dominant dipolar contribution
$-4m^2/r_c^3$ and additional terms of higher order in $r_c^{-1}$.

The direct electrostatic energy $u_{el}$ is a sum of pair-additive
interactions with the other molecules. It closely resembles the
expression from standard electrostatics, with the interaction $1/r$
being replaced by $\varphi({\bf r})$. Based on this resemblance,
$u_{el}$ could form a possible choice in the calculation of energies
of charging $\Delta\! u$. The self-term $u_s$ is a consequence of the
difference of the effective interaction $\varphi({\bf r})$ and
$1/r$. In the following, we will consistently include the self-term
$u_s$ in the calculation of electrostatic energies. Taking only
$u_{el}$ would be correct for an isolated system, but not when
periodic boundary conditions are taken seriously. The direct
electrostatic energy $u_{el}$ stems from the interactions with the
other particles in the box (and, in the case of Ewald summation, their
periodic images). The self-term $u_s(\lambda)$ is the difference of
the self-interactions with $\varphi$ as effective potential and the
vacuum self-interactions with a $1/r$ potential (which cancel the
infinities of the former). In the case of Ewald summation, $u_s$ has
physical significance as it accounts for the interactions with the
images of the water molecule with charges $(1-\lambda)q_\alpha$. In
the context of GRF electrostatics, $u_s$ is introduced by analogy.

In a previous simulation study, Rick and Berne \cite{Rick:94} also
used Ewald summation to calculate the free energy of water but with a
potential corresponding to vacuum boundary conditions. Their effective
potential for $u_{el}$ calculations differs by $-2\pi r^2/3V$ (and a
constant) from the Ewald potential used here. Correspondingly, the
self-interaction of eq~\ref{eq:PsiEW} then reduces to a constant and
the self-term of Rick and Berne for neutral molecules is zero.

\section{Computer simulations}
We study bulk water ($\lambda=0$) and a system comprising one
uncharged and $N-1$ unmodified water molecules ($\lambda=1$). We use
the SPC model of water.\cite{Berendsen:81} Configuration-space
averages are performed using the Metropolis Monte Carlo
method.\cite{Allen:87,Metropolis:53} The systems are studied in the
canonical ({\em NVT}) ensemble. The number density is $\rho =
33.33$~nm$^{-3}$. The temperature is kept at $T=298$~K. The Monte
Carlo move widths is chosen to give an approximate acceptance ratio of
0.5. A cubic box under periodic boundary conditions is used for the
simulations. The system size is varied between $N=256$, 64, and 32
water molecules to study finite-size effects.

The long-range electrostatic interactions are treated using Ewald
summation.\cite{Allen:87,Ewald:21,deLeeuw:80:a} The real-space
screening factor is set to $\eta=5.6/L$ ($L$ is the length of the
box). The real-space interactions are truncated at $L/2$. The
Fourier-space sum is spherically truncated using ${\bf k}$ vectors
with $|{\bf k}|^2\leq 38(2\pi/L)^2$, resulting in $2\times 510$ ${\bf
k}$ vectors considered.\cite{Hummer:95:a} The background dielectric
constant is corrected from infinity to $\epsilon_{RF}=65$ by adding a
term $2\pi{\bf M}^2/(2\epsilon_{RF}+1)V$ to the potential energy,
where ${\bf M}$ is the total dipole moment of the simulation box.

We also perform simulations using the GRF interaction. A cutoff of
$r_c=0.9$~nm ($N=256$) and $r_c=L/2$ ($N=64$) is used. Again, a
correction term for a finite dielectric background is added to the
total energy. In all simulations, periodic boundary conditions are
applied with respect to atomic sites. As starting structures we use
randomly oriented and positioned water molecules or final structures
or previous runs. Before averaging, the systems are extensively
equilibrated.

To calculate the electrostatic energy of charging an uncharged water
molecule, systems are studied comprising $N-1$ SPC water molecules and
one uncharged Lennard-Jones sphere with water parameters. The number
density is $\rho=33.33$~nm$^{-3}$. At regular intervals, the
electrostatic energy is calculated for 50 random orientations of a
fictitious charged molecule with the oxygen position identical to the
uncharged Lennard-Jones particle.

\section{Results and discussion}
We will first present results for the statistical distribution of
$u_{el}$. Figure~\ref{fig:hist} shows histograms of $u_{el}$ for the
charged and uncharged state calculated using $N=256$ particles and
Ewald summation. The histograms are shown on a logarithmic scale so
that Gaussians appear as parabolas. Also shown are Gaussian
distributions with identical mean and variance. The distribution in
the charged case ($\lambda=0$) is approximately Gaussian, but more
centered. It decays faster than Gaussian and has a negative
kurtosis. The distribution in the uncharged case ($\lambda=1$) on the
other hand deviates more strongly from a Gaussian form. It is skewed,
has weakly decaying tails and, correspondingly, a positive kurtosis.

Table~\ref{tab:cum} lists the cumulants $C_{n,\lambda}$ of $u_{el}$
for simulations with $\lambda=0$ and 1. Results are shown for
Ewald-summation and GRF electrostatics and for system sizes of $N=32$,
64, and 256 molecules. The results for small system sizes are included
to study the dependence of the cumulants on electrostatic interaction
and system-size, as we are ultimately interested in the thermodynamic
limit.
The consideration of the self-term greatly reduces the
variation for the averages $C_{1,\lambda=0}$. The uncorrected
Ewald-summation results for $N=32$, 64, and 256 range between $-96.69$
and $-98.00$~kJ~mol$^{-1}$. With the self-term added, the range is
$-98.05$ to $-98.17$~kJ~mol$^{-1}$, compared to a typical statistical
error of 0.1~kJ~mol$^{-1}$.\cite{errors} The GRF self-term is
significantly larger than that of Ewald summation. The corrected value
$C_{1,\lambda=0}'$ of the GRF simulation with $N=256$ agrees with the
Ewald-summation result within the statistical errors. The corrected
$N=64$ result for $C_{1,\lambda=0}'$ is somewhat too negative, but in
much better agreement with the Ewald-summation data (1 versus
10~kJ~mol$^{-1}$ difference).

On the other hand, the variance data $C_{2,\lambda}$ from
Ewald-summation simulations do not show a clear improvement by
addition of the self-term. However, this is not conclusive since the
corrections are of the size of the statistical error or smaller. The
improvement by addition of the self-term is more evident in the GRF
variances. The higher cumulants $C_{3,\lambda=0}$ and
$C_{4,\lambda=0}$ show less system-size dependence. Results for
different system sizes and electrostatics agree within statistical
errors. A possible exception is that the GRF values for
$C_{3,\lambda=0}$ are too small. Overall, the first four cumulants of
the charged state ($\lambda=0$) can be calculated accurately for
systems with as few as $N=64$ (and possibly $N=32$) water molecules
using Ewald summation if the finite-size corrections are applied.

We use the cumulants $C'_{1,\lambda}$, $C'_{2,\lambda}$,
$C_{3,\lambda}$, and $C_{4,\lambda}$ for the charged ($\lambda=0$) and
uncharged ($\lambda=1$) case of the $N=256$ simulations to calculate
the chemical-potential difference $\Delta\!\mu^{ex}$ of uncharged and
charged water. The cumulants $C_{4,\lambda}$ for $\lambda=0$ and 1
(see Table~\ref{tab:cum}) differ significantly. Therefore, we have to
use fitting polynomials of order five or higher in the $\chi^2$ fit
eq~\ref{eq:chi2}. Results for $\Delta\!\mu^{ex}$ are listed in
Table~\ref{tab:mu}. The polynomial $p_8$ of order 8 is interpolating,
i.e., $p_8$ fits the data exactly. The results for polynomials
$p_5$ to $p_8$ vary between $\Delta\!\mu^{ex} = 35.13$ and
35.80~kJ~mol$^{-1}$, or by about 2~{\%}. The Ewald and GRF results for
like orders differ by less than 0.09~kJ~mol$^{-1}$ (0.25~{\%}). The
interpolating polynomials $p_8$ yield 35.60 (Ewald) and
35.63~kJ~mol$^{-1}$ (GRF) for $\Delta\!\mu^{ex}$. From block averages,
we estimate the statistical errors of the $p_8$ data for
$\Delta\!\mu^{ex}$ to be 0.15~kJ~mol$^{-1}$.

We will now compare the results from polynomial fits with those
obtained from Bennett's method of overlapping
histograms.\cite{Bennett:76} Figure~\ref{fig:bennett} shows the ratio
$\ln(p_1/p_0)$ of the probabilities as a function of $\Delta\! u =
u(\lambda_1)-u(\lambda_0)$, as described above. The distributions
$p_1$ and $p_0$ are calculated from histograms of $u_{el}$ using a bin
widths of 0.1~kJ~mol$^{-1}$. Self-interactions are added and the sign
is inverted according to eq~\ref{eq:dubenn}. In the overlap region of
the distributions, we indeed find the expected linear behavior
$\ln(p_1/p_0)=\beta(\Delta\!\mu^{ex} -\Delta\! u)$. We fit constants
$\Delta\!\mu^{ex}=35.63$ and 35.60~kJ~mol$^{-1}$ to the
Ewald-summation and GRF data for $20\leq \Delta\! u \leq
50$~kJ~mol$^{-1}$, respectively. The $\Delta\!\mu^{ex}$ values are in
excellent agreement with $\Delta\!\mu^{ex}=35.60$ (Ewald) and
35.63~kJ~mol$^{-1}$ (GRF) calculated from the polynomials $p_8$
interpolating the derivative data.

Also included in Figure~\ref{fig:bennett} are the results of Gaussian
approximations to the probability densities. However, as expected from
the significant deviations from Gaussian behavior found in
Figure~\ref{fig:hist}, we do not observe a linear regime. This
provides evidence for the failure of a simple Gaussian picture of the
fluctuation statistics for the hydration of water. Assumption of
Gaussian behavior (or, correspondingly, quadratic dependence on the
coupling parameter) results in inaccurate estimates of the
chemical-potential difference $\Delta\!\mu^{ex}$.

Bennett's acceptance-ratio method\cite{Bennett:76} yields 35.59
(Ewald) and 35.60~kJ~mol$^{-1}$ (GRF) for $\Delta\!\mu^{ex}$. These
values are calculated by searching for the intersection of the
Fermi-function averages for $\lambda=0$ and 1 in eq~\ref{eq:acc}. The
corresponding value of $c$ is identified with the difference in the
chemical potential, $\beta \Delta\!\mu^{ex} = c$. The averages are
calculated from the tabulated histogram data for $\Delta\! u$. The
acceptance-ratio values of $\Delta\!\mu^{ex}$ are again in excellent
agreement with those calculated using the interpolating polynomial
$p_8$.

Rick and Berne\cite{Rick:94} calculated $\Delta\!\mu^{ex}$ by
evaluating $C_{1,\lambda}$ at different charge states
$-0.22\leq\lambda\leq 1$ and combining the data by numerical
integration. This is essentially a variant of the method we are using
in this work. However, inclusion of accurately sampled cumulants of
higher order---as is done here---should decrease the number of $\lambda$
points required. From eleven molecular dynamics simulations of 40~ps
duration with $N=512$ SPC water molecules, Rick and Berne find
$\Delta\!\mu^{ex}=35.15\pm 2.1$~kJ~mol$^{-1}$, in agreement with our
data.

Figure~\ref{fig:mulam} shows the chemical potential $\Delta\!\mu^{ex}$
as a function of the charge coupling parameter $\lambda$. Results are
shown for the interpolating polynomials $p_8$ of the Ewald-summation
and GRF data, which are practically indistinguishable. Also shown are
the data of Rick and Berne,\cite{Rick:94} which are found to agree
with the $p_8$ curves. We find agreement also in the region
$\lambda<0$, where the polynomials $p_8$ are extrapolating the data at
$\lambda=0$ and 1.

The parabolic $\Delta\!\mu^{ex}$ curves obtained by assuming
Gaussian-fluctuation statistics are included in
Figure~\ref{fig:mulam}. The quadratic expansions around both the
charged and uncharged state are accurate only for small perturbations
$|\Delta\!\lambda|<0.3-0.4$. For larger perturbations, the parabolas
show large deviations from the fitted polynomials and the data of Rick
and Berne. This provides further evidence for the failure in the
large-perturbation regime of simple models, which represent the
chemical potential as quadratic functions of the coupling parameter
based on the assumption of Gaussian-fluctuation statistics.

Also shown in Figure~\ref{fig:mulam} are the results of direct
calculations using eq~\ref{eq:dF}, which can be rewritten for the
case of uncharging water as
\begin{eqnarray}
\mu^{ex}(\lambda_1)-\mu^{ex}(\lambda_0)&=&-k_B T\ln
\left\langle \exp [ \beta ( \lambda_1 - \lambda_0 ) u_{el}(0) ]
\right\rangle_{\lambda_0} + \Delta\!\lambda ( \Delta\!\lambda +
2\lambda_0 - 2 ) u_s(0)~.
\label{eq:direct}
\end{eqnarray}
Correspondingly, the direct calculation determines the logarithm of
the characteristic function of the distribution of electrostatic
energies $u_{el}(0)$. We can see that the exponential average for both
$\lambda_0=0$ and 1 works well for small perturbations
$|\lambda_1-\lambda_0|<0.5$. For larger perturbations, the direct
method fails. Expansions around the charged and uncharged state
predict values for the difference in chemical potential which are too
large by about 2.8 and too small by about 6.8~kJ~mol$^{-1}$,
respectively.

However, by fitting polynomials to the derivatives at the two ends
$\lambda=0$ and 1 of the interval, we essentially attempt to connect
the two curves smoothly near $\lambda=0.5$. This explains the success
of the method based on fluctuation statistics: We can accurately
combine information of two widely separated states by smoothly
connecting the curves from direct averages near $\lambda=0.5$, where
they are still accurate. Using higher-order cumulants which can be
sampled accurately in this case since they are significantly different
from zero allows us to construct a non-trivial form of the chemical
potential as a function of charge. This charging function deviates
strongly from a simple quadratic form, in agreement with previous
observations by Rick and Berne\cite{Rick:94} and with large values of
the kurtosis $C_4/C_2^2$ of the $u_{el}$ distributions.

The idea of smoothly connecting the curves from direct calculations
can be turned into an algorithm. We shift the $\lambda=1$ curve
$\gamma_1 (\lambda) = - k_B T \ln \langle \exp[ - \beta u(\lambda) + \beta
u(1) ] \rangle_1$ vertically by $\Delta\!\mu^{ex}$ and determine the
intersection with the $\lambda=0$ curve $\gamma_0 (\lambda) = - k_B T
\ln \langle \exp[ - \beta u(\lambda) + \beta u(0) ] \rangle_0$. We then
calculate the first derivatives of the two curves at the intersection
from polynomial interpolation. The absolute value of the difference
$\Delta=\gamma_1' (\bar{\lambda}) - \gamma_0' (\bar{\lambda})$ between
the two derivatives is minimized with respect to the vertical shift,
\begin{eqnarray}
\min_{\Delta\!\mu^{ex}} \left| \gamma_1' (\bar{\lambda}) -
\gamma_0' (\bar{\lambda}) \right|~,
\end{eqnarray}
where $\bar{\lambda}$ is the solution of $\gamma_0(\bar{\lambda}) =
\gamma_1(\bar{\lambda}) + \Delta\!\mu^{ex}$. $|\Delta|$ is
plotted in Figure~\ref{fig:dmudl} as a function of the
chemical-potential difference $\Delta\!\mu^{ex}$ between the charged
and uncharged state. The difference in the derivatives shows a
distinct minimum for $\Delta\!\mu^{ex}$ between 35.7 and
35.8~kJ~mol$^{-1}$. This result is in excellent agreement with the
previously calculated numbers, deviating by less than 0.1~{\%}. This
simple, geometrical analysis can therefore complement the more
elaborate methods and provide accurate results for the chemical
potential.

We can also use this analysis to choose an optimal $\lambda$ value in
the interval $[\lambda_0,\lambda_1]$ to improve the accuracy of the
results. From the $|\Delta|$ dependence on the $\bar{\lambda}$-value
of intersection of the two curves $\gamma_0(\bar{\lambda})$ and
$\gamma_1(\bar{\lambda}) + \Delta\!\mu^{ex}$, we find that the
smoothest connection occurs for $\bar{\lambda}\approx 0.45-0.5$. For
the purpose of calculating free energies, it would be interesting to
perform a single simulation at $\lambda=0.5$ (i.e., for a water
molecule carrying 0.5 times the full charges). Judging from the
previous results, a simulation at $\lambda=0.5$ should produce
sufficient information to calculate the chemical potential on the
whole interval $0\leq\lambda\leq 1$. However, in this work we restrict
our analysis to the physically more relevant systems of bulk water
($\lambda=0$) and an uncharged solute in water ($\lambda=1$).

Another observation of this study is the importance of the inclusion
of self-terms as approximate corrections for finite-size
effects. Included in Table~\ref{tab:mu} are the $\Delta\!\mu^{ex}$
values calculated from polynomial fits to the data of the $N=64$
simulations using Ewald summation. The values of the $N=64$ and
$N=256$ systems agree closely. The difference of the $p_8$ data is
0.04~kJ~mol$^{-1}$ compared to estimated statistical errors of
0.15~kJ~mol$^{-1}$. On the other hand, the $p_8$ values for the
uncorrected cumulant data (i.e., without self-terms) are 35.22
($N=64$) and 35.52~kJ~mol$^{-1}$ ($N=256$), differing by
0.3~kJ~mol$^{-1}$. Addition of the self-term results in a reduced
system-size dependence.

The improvement is even more evident when we compare results from
simulations using different electrostatic interactions. The difference
in chemical potential $\Delta\!\mu^{ex}$ of charged and uncharged
state calculated from the polynomial fit $p_8$ and Bennett's
overlapping-histogram and acceptance-ratio methods agree within
0.04~kJ~mol$^{-1}$ for GRF and Ewald-summation data if self-terms are
added. Without the self-terms they would differ by about
1.7~kJ~mol$^{-1}$. In addition, also the chemical potential as a
function of the charge shown in Figure~\ref{fig:mulam} is identical in
the range of $\lambda$ considered. This shows that even with less
sophisticated methods for the electrostatics such as GRF compared to
Ewald summation it is possible to calculate charge-related free
energies accurately if only self-terms are considered consistently.

\section{Conclusions}
Using fluctuation statistics of two equilibrium simulations (bulk
water and water hydrating an uncharged Lennard-Jones particle) we are
able to calculate accurately the chemical potential of SPC water as a
function of its charge. Bennett's overlapping-histogram and
acceptance-ratio methods\cite{Bennett:76} and a geometrical method
based on smoothly connecting the curves of direct, exponential
averages at the two states give identical results for the
chemical-potential difference between charged and uncharged water. Our
results agree with those of Rick and Berne\cite{Rick:94} calculated
from thermodynamic integration using eleven states with respect to
both the chemical-potential difference and the dependence on the
charge.

We find significant deviations from a quadratic dependence of the
chemical potential on the charge coupling parameter. This has
important implications. It shows that even for the fairly simple
system of water in water second-order perturbation or, equivalently,
assumption of Gaussian-fluctuation statistics, allow only crude
descriptions of the actual thermodynamics. This also affects the
potential usefulness of linear continuum models of
electrostatics,\cite{Pratt:94:a,Rick:94,Davis:90:a,Honig:93,%
Tawa:94:a,Tawa:95:a} which by design do not go beyond a quadratic
behavior. However, when calculating free-energy differences, continuum
models usually compare approximate representations of physical states
rather than performing an {\em expansion} around a single state. This
can explain their success of giving at least approximately correct
free-energy values even in the presence of strongly non-linear
dependencies in a corresponding system of atomic
resolution.\cite{Rick:94}

In addition to the failure of the simple Gaussian model, expansions
around only a single state using higher-order cumulants are expected
to fail.\cite{Smith:94} Increasing the order of the perturbation
requires accurate information about the poorly sampled tails of
distributions. This merely reflects the difficulty to {\em
extrapolate} to states that strongly differ in their structure and
fluctuation statistics. {\em Interpolation} on the other hand is in
general a much simpler task and allows more accurate predictions. In
this work we effectively combine the information of two states to
derive a polynomial expression for the chemical potential.

An important point concerns the treatment of electrostatic
interactions in computer simulations of systems under periodic
boundary conditions. With respect to system-size dependence and
electrostatic model, we obtain consistent results using Ewald
summation and a generalized reaction-field model. But consistency is
only achieved if self-interactions are included. These
self-interactions arise naturally when effective potentials are used
for the Coulomb interactions. We made similar observations in previous
studies of the chemical potential of ions.\cite{Hummer:95:c,Hummer:93}
Adding self-terms to the energies acts as an effective correction
for effects of a finite system size. Neglecting them can result in
significant deviations for small systems of a few hundred particles.

\section*{Acknowledgments}
The authors wish to thank M. Neumann and D. M. Soumpasis for many
stimulating discussions. This work was supported by the Department of
Energy (U.S.).

\begin{figure}
\caption{Probability density of the electrostatic energy $u_{el}$ of a
water molecule with full charge ($\lambda=0$; centered near
$u_{el}=-100$~kJ~mol$^{-1}$) and with zero charge, where fictitious
charges are turned on ($\lambda=1$; centered near $u_{el}=0$). The
probabilities are shown on a logarithmic scale such that Gaussians
appear as parabolas. Also shown are Gaussian distributions (dot-dashed
lines) with mean and variance equal to the calculated
distributions. The distributions are calculated from Monte Carlo
simulations using $N=256$ particles with Ewald summation. The curves
do not contain corrections for self-interactions.\label{fig:hist}}
\end{figure}

\begin{figure}
\caption{Bennett's method of overlapping
histograms.\protect\cite{Bennett:76} The ratio $\ln(p_1/p_0)$ of the
probabilities is shown with crosses as a function of $\Delta\! u =
u(\lambda_1=1)-u(\lambda_0=0)$. $p_1$ and $p_0$ are approximated by
histogram values calculated from Ewald-summation and GRF simulations
using $N=256$ particles. The GRF data are shifted vertically by
$-10$. Also shown are lines with slope $-\beta$ fitting the data. The
dot-dashed lines represent the results of a Gaussian approximation
(see Figure~\protect\ref{fig:hist}) with mean and variance taken from
the GRF and Ewald-summation data. \label{fig:bennett}}
\end{figure}

\begin{figure}
\caption{The chemical potential $\Delta\!\mu^{ex}$ of water as a
function of its charge. $\lambda=0$ and 1 correspond to the fully
charged and uncharged state, respectively. Solid lines are polynomials
of order 8 fitted to the cumulants $C'_{1,\lambda}$, $C'_{2,\lambda}$,
$C_{3,\lambda}$, and $C_{4,\lambda}$ for
$\lambda=0$ and 1. The curves for Ewald-summation and GRF data are
practically indistinguishable. Shown as symbols with error bars are
the data of Rick and Berne taken from Figure~2 of
ref~\protect\onlinecite{Rick:94}. Also shown are results from direct
calculations using eq~\protect\ref{eq:direct}. The expansion around
charged and uncharged state $\lambda=0$ and $\lambda=1$ are shown with
dashed and dot-dashed lines, respectively, in both cases using the
data of Ewald-summation simulations. The dotted lines represent
quadratic expansions around $\lambda=0$ and 1 using mean and variance
of the Ewald-summation data and assuming Gaussian
statistics. \label{fig:mulam}}
\end{figure}

\begin{figure}
\caption{Smooth connection of the chemical-potential curves. Two curves
for the chemical potential are calculated from direct, exponential
averages eq~\protect\ref{eq:direct} for the charged and uncharged
state. One of the curves is shifted vertically and the intersection of
the two curves is calculated. Shown is the absolute value of the
difference $\Delta$ of the first derivatives at the intersection as a
function of the corresponding difference in chemical potential
$\Delta\!\mu^{ex}$. Where $|\Delta|$ reaches a minimum, the two curves
can be connected most smoothly. This determines an estimated value for
the chemical-potential difference between charged and uncharged water
$\Delta\!\mu^{ex} = 35.7 - 35.8$~kJ~mol$^{-1}$. \label{fig:dmudl}}
\end{figure}

\begin{table}
\caption{Expansion-coefficients $S$ and $A_l$ of the Ewald potential
$\varphi_{EW}({\bf r})$ in terms of kubic-harmonic
functions\protect\cite{Hummer:93} in the convention of Adams and
Dubey.\protect\cite{Adams:87} The constant $S$ has been adjusted from
its exact value such that the numerical scheme gives a vanishing
average potential of a point charge in a cube.\label{tab:harm}}
\begin{tabular}{cc}
$S      $&$ -2.8373002368 $\\
$A_2    $&$ 2\pi/3        $\\
$A_4    $&$ 7.718196      $\\
$A_6    $&$ 20.657378665  $\\
$A_8    $&$ 86.85346475   $\\
$A_{10} $&$ 179.631024892 $
\end{tabular}
\end{table}

\begin{table}
\caption{Cumulants of the $u_{el}$ distributions calculated from Monte
Carlo simulations using $N$ particles. The cumulants were averaged
over $P$ passes, where one pass consists of $N$ attempted moves (i.e.,
one attempted Monte Carlo move per particle). Coulomb denotes the
treatment of electrostatic interactions. The fully charged and
uncharged water molecules correspond to $\lambda=0$ and
$\lambda=1$. $C_k$ denotes the $k$-th cumulant of the distribution
[measured in (kJ~mol$^{-1}$)$^k$]. The cumulants with corrections for
the self-interaction are listed as $C_k'$.
\label{tab:cum}}
\begin{tabular}{cccccccccc}
$N$ & Coulomb & $P/1000$ & $\lambda$ & $C_1$ & $C_1'$ & $C_2$ & $C_2'$
& $C_3$ & $C_4$ \\ \hline
$256 $& EW  &$ 280 $&$ 0 $&$ -98.00(10) $&$ -98.17(10) $&$ 439.5(2.0)
 $&$ 440.0(2.0) $&$  120(50) $&$ -25400(1000) $\\
$ 64 $& EW  &$ 860 $&$ 0 $&$ -97.47(10) $&$ -98.15(10) $&$ 439.7(2.0)
 $&$ 441.4(2.0) $&$  102(60) $&$ -24900(1500) $\\
$ 32 $& EW  &$ 860 $&$ 0 $&$ -96.69(15) $&$ -98.05(15) $&$ 443.4(3.0)
 $&$ 446.8(3.0) $&$  100(90) $&$ -27200(2500) $\\
$256 $& GRF &$ 460 $&$ 0 $&$ -94.73(10) $&$ -98.22(10) $&$ 434.5(1.5)
 $&$ 443.1(1.5) $&$  62(50)  $&$ -24400(1000) $\\
$ 64 $& GRF &$ 900 $&$ 0 $&$ -88.36(10) $&$ -99.05(10) $&$ 418.3(1.5)
 $&$ 444.7(1.5) $&$  21(40)  $&$ -23650(1000) $\\
$256 $& EW  &$ 280 $&$ 1 $&$ -0.006(5)  $&$ -0.006(5)  $&$ 110.0(2.0)
 $&$ 110.4(2.0) $&$  261(20) $&$ 13000(1000)  $\\
$ 64 $& EW  &$1000 $&$ 1 $&$ -0.007(5)  $&$ -0.007(5)  $&$ 109.5(3.0)
 $&$ 111.2(3.0) $&$  263(15) $&$ 12500(1000)  $\\
$256 $& GRF &$ 600 $&$ 1 $&$ 0.16(2)    $&$ 0.16(2)    $&$ 106.4(2.0)
 $&$ 115.1(2.0) $&$  277(20) $&$ 12450(1000)  $
\end{tabular}
\end{table}

\begin{table}
\caption{Difference in chemical potential $\Delta\!\mu^{ex}$
of an uncharged and charged water molecule in water. Results from
$\chi^2$ fits of polynomials $p_n$ of order $n$ to the derivative data
of the $N=256$ simulations using Ewald-summation (EW-256) and GRF
interactions. Also included are results for the $N=64$ simulations
using Ewald summation (EW-64). $\Delta\!\mu^{ex}$ is listed in
kJ~mol$^{-1}$. The statistical errors of the $p_8$ data are estimated
to be 0.15~kJ~mol. \label{tab:mu}}
\begin{tabular}{cccc}
      & EW-256  & EW-64 & GRF   \\ \hline
$p_5$ & 35.46   & 35.23 & 35.45 \\
$p_6$ & 35.13   & 35.05 & 35.22 \\
$p_7$ & 35.76   & 35.71 & 35.80 \\
$p_8$ & 35.60   & 35.56 & 35.63
\end{tabular}
\end{table}

\begin{thebibliography}{10}

\bibitem{Frenkel:86}
Frenkel, D. In {\em Molecular Dynamics Simulations of Statistical
  Mechanical Systems. Proceedings of the Enrico Fermi Summer School,
  Varenna, 1985}; Ciccotti, G.; Hoover, W.~G., Eds.; North-Holland:
  Amsterdam, 1986, p~151.

\bibitem{Pratt:94:a}
Pratt, L.~R.; Hummer, G.; Garc\'{\i}a, A.~E. {\em Biophys. Chem.} {\bf
  1994}, {\em 51}, 147.

\bibitem{Hummer:95:c}
Hummer, G.; Pratt, L.~R.; {Garc\'{\i}a}, A.~E. {\em J. Chem. Phys.}
  {\bf 1996} (in press).

\bibitem{Born:20}
Born, M. {\em Z. Phys.} {\bf 1920}, {\em 1}, 45.

\bibitem{Levy:91}
Levy, R.~M.; Belhadj, M.; Kitchen, D.~B. {\em J. Chem. Phys.} {\bf
  1991}, {\em 95}, 3627.

\bibitem{Figueirido:94}
Figueirido, F.; Del Buono, G.~S.; Levy, R.~M. {\em Biophys. Chem.} {\bf
  1994}, {\em 51},  235.

\bibitem{Widom:82}
Widom, B. {\em J. Phys. Chem.} {\bf 1982}, {\em 86}, 869.

\bibitem{Bennett:76}
Bennett, C.~H. {\em J. Comput. Phys.} {\bf 1976}, {\em 22}, 245.

\bibitem{Rick:94}
Rick, S.~W.; Berne, B.~J. {\em J. Am. Chem. Soc.} {\bf 1994}, {\em
116}, 3949.

\bibitem{Hummer:93}
Hummer, G.; Soumpasis, D.~M. {\em J. Chem. Phys.} {\bf 1993}, {\em
98}, 581.

\bibitem{Allen:87}
Allen, M.~P.; Tildesley, D.~J.  {\em Computer Simulation of Liquids};
  Clarendon Press: Oxford, UK, 1987.

\bibitem{Levesque:92}
Levesque, D.; Weis, J.~J. In {\em The {Monte} {Carlo} Method in
  Condensed Matter Physics}; Binder, K., Ed.; Springer: Berlin, 1992,
  p~121.

\bibitem{Stuart:87}
Stuart, A.~S.; Ord, J.~K.  {\em {Kendall's} Advanced Theory of
  Statistics}; Vol. 1; 5th ed; Oxford University Press: Oxford, UK,
  1987.

\bibitem{Marcinkiewicz:39}
Marcinkiewicz, J. {\em Math. Zeitschr.} {\bf 1939}, {\em 44},  612.

\bibitem{Gubbins:83}
Gubbins, K.~E.; Shing, K.~S.; Streett, W.~B. {\em J. Phys. Chem.} {\bf
  1983}, {\em 87}, 4573.

\bibitem{Ewald:21}
Ewald, P.~P. {\em Ann. Phys.} {\bf 1921}, {\em 64}, 253.

\bibitem{deLeeuw:80:a}
de~Leeuw, S.~W.; Perram, J.~W.; Smith, E.~R. {\em
  Proc. R. Soc. Lond. A} {\bf 1980}, {\em 373}, 27.

\bibitem{Slattery:80}
Slattery, W.~L.; Doolen, G.~D.; DeWitt, H.~E. {\em Phys. Rev.} {\bf
  1980}, {\em A21}, 2087.

\bibitem{Adams:87}
Adams, D.~J.; Dubey, G.~S. {\em J. Comput. Phys.} {\bf 1987}, {\em
72}, 156.

\bibitem{vdLage:47}
{von der Lage}, F.~C.; Bethe, H.~A. {\em Phys. Rev.} {\bf 1947}, {\em
  71}, 612.

\bibitem{Hummer:94:a}
Hummer, G.; Soumpasis, D.~M. {\em Phys. Rev. E} {\bf 1994}, {\em 49},
591.

\bibitem{Hummer:94:e}
Hummer, G.; Soumpasis, D.~M.; Neumann, M. {\em J. Phys.:
  Condens. Matt.} {\bf 1994}, {\em 23A}, A141.

\bibitem{Berendsen:81}
Berendsen, H. J.~C.; Postma, J. P.~M.; {van Gunsteren}, W.~F.;
  Hermans, J. In {\em Intermolecular Forces: Proceedings of the 14th
  Jerusalem Symposium on Quantum Chemistry and Biochemistry}; Pullman,
  B., Ed.; Reidel: Dordrecht, Holland, 1981, p~331.

\bibitem{Metropolis:53}
Metropolis, N.; Rosenbluth, A.~W.; Rosenbluth, M.~N.; Teller, A.~H.;
  Teller, E.  {\em J. Chem. Phys.} {\bf 1953}, {\em 21}, 1087.

\bibitem{Hummer:95:a}
Hummer, G. {\em Chem. Phys. Lett.} {\bf 1995}, {\em 235}, 297.

\bibitem{errors}
Statistical errors were determined from block averages. The listed
error estimates were rounded up and correspond to typically 1.5
estimated standard deviations as calculated from blocks of 20,000 or
40,000 Monte Carlo passes.

\bibitem{Davis:90:a}
Davis, M.~E.; McCammon, J.~A. {\em Chem. Rev.} {\bf 1990}, {\em 90},
509.

\bibitem{Honig:93}
Honig, B.; Sharp, K.; Yang, A.-S. {\em J. Phys. Chem.} {\bf 1993},
  {\em 97}, 1101.

\bibitem{Tawa:94:a}
Tawa, G.~J.; Pratt, L.~R. In {\em Structure and Reactivity in Aqueous
  Solution: Characterization of Chemical and Biological Systems. {ACS}
  Symposium Series}; Cramer, C.~J.; Truhlar, D.~G., Eds.; ACS:
  Washington DC, 1994, Vol.~568, p~60.

\bibitem{Tawa:95:a}
Tawa, G.~J.; Pratt, L.~R. {\em J. Am. Chem. Soc.} {\bf 1995}, {\em
117}, 1625.

\bibitem{Smith:94}
Smith, P.~E.; {van Gunsteren}, W.~F. {\em J. Chem. Phys.} {\bf 1994},
  {\em 100}, 577.

\end{thebibliography}
\end{document}